\documentclass[11pt,twoside]{article}
\usepackage{asp2010}

\resetcounters

\begin{document}

\title{The DASCH Data Processing Pipeline and Multiple Exposure Plate Processing.}
\author{Edward~Los$^1$, Jonathan~Grindlay$^1$, Sumin~Tang$^1$, Mathieu~Servillat$^1$, and Silas~Laycock$^2$
\affil{$^1$Harvard College Observatory}
\affil{$^2$Department of Physics, University of Massachusetts at Lowell}}

\begin{abstract}
Digital Access to a Sky Century @ Harvard (DASCH) is a project to digitize the collection of approximately 525,000 astronomical plates held at the Harvard College Observatory. This paper presents an overview of the DASCH data processing pipeline, with special emphasis on the processing of multiple-exposure plates.  Such plates extended the dynamic range of photograph emulsions and improved photometric accuracy by minimizing variations in plate development procedures.  Two approaches are explored in this paper:  The repetitive use of astrometry.net \citep{lang2009} and local correlation searches.  Both procedures have yielded additional quality control checks useful to the pipeline.

\end{abstract}

\section{Introduction}

The Harvard College Observatory plate collection consists of approximately 525,000 photographs produced by over 80 telescopes spanning over 100 years from about 1885 to 1992.  The goal of the Digital Access to a Sky Century @ Harvard\footnote{see http://hea-www.harvard.edu/DASCH/.}  (DASCH) project \citep{grindlay2009} is to digitize this entire collection and provide photometry measurements for all objects.  To date, we have digitized over 10,000 plates and extracted an average of 80,000 objects per plate.  The analysis of this digitized data presents a number of challenges which are no longer encountered with modern CCD photographic techniques.  One of these challenges is the presence of many plates which have multiple instances of the same objects. \footnote{see http://hea-www.harvard.edu/DASCH/papers/P033f.pdf for a full-length version of this paper.} 
\subsection{Types of Multiple Exposure Plates}

\begin{figure}
\plotone{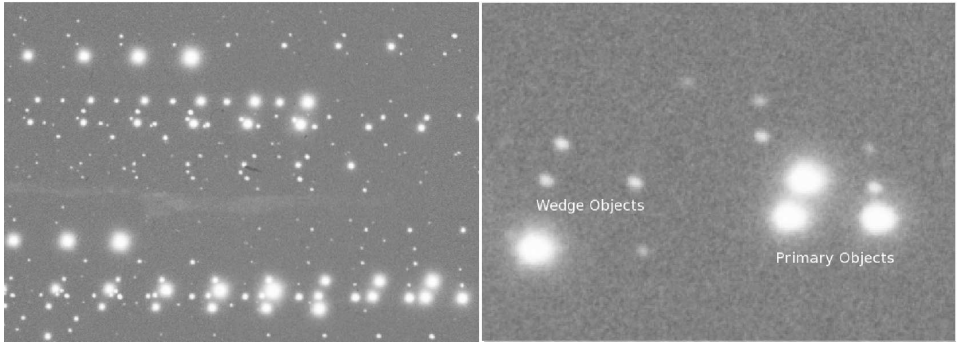}
\caption{Left: Plate mc05077 showing multiple exposures of the M44 field. Nine exposures were taken with exposure times decreasing by 50\% for each successive exposure. Right: Plate i31090 showing examples of Pickering Wedge objects.\label{f1}}
\end{figure}

Plates with multiple exposures were produced to extend the limited dynamic range of photographic emulsions, to account for variations in development procedures, and to include known standards such as the Harvard North Polar Sequence.  Figure~\ref{f1} shows examples successive exposures of the same field and of ghost objects generated by  Pickering or Racine Wedge \citep{leavitt1917}.  A third technique for producing ghost objects made use of coarse gratings \citep{king1930}.
\subsection{Number of Multiple Exposure Plates}

Now that 217,000 of the 525,000 logbook entries have been transcribed and entered into a MySQL database, a simple query shows that 5,842 or 2.6\% of the transcribed plates have multiple exposures.  Unfortunately, no mention of the use of Pickering Wedge plates has been found in the logbooks.  Logbook transcriptions suggest that less than 100 plates used coarse gratings, although the total may be as high as 8000 because of issues with the logbook plate classification system.

\section{DASCH pipeline}

An overview of the DASCH digitizer and pipeline appears below.  The two key steps involved in the processing of multiple exposure plates are the ``Pickering Wedge Filter'' and the ``Multiple Exposure Loop''.  

\subsection{Plate Preparation}

Before any scanning can occur, the relevant entries in the logbook must be transcribed and entered into the MySQL scanner database.  Both plate jackets and plates with ink annotations are photographed with a Nikon D200 camera.  All ink annotations on the reverse side of the plate from the emulsion must then be cleaned to avoid confusion with astronomical objects.

\subsection{Mosaic Generation, WCS fitting, and Source Extraction}

The digitizer \citep{simcoe2006} generates 60 tiles to cover a typical 20 x 25 cm plate with 10 x 6 images using half-steps in width to assure two exposures for every plate object.  The mosaicing process registers and combines these tiles with flatfield tiles to produce a single mosaic of approximately 780 megapixels.

The SExtractor program \citep{bertin1996} next generates object lists.  The WCS fitting procedure first begins with astrometry.net \citep{lang2009} and moves to successively accurate fits using WCStools \citep{mink1999} and then a polynomial fit.   A companion paper \citep{servillat2010} describes this procedure in greater detail.

\subsection{Pickering Wedge Filter}

The Pickering Wedge Filter is used to flag ghost objects by performing a spatial correlation within a limited region around bright stars. The procedural steps are as follows: (1) Select the 300 brightest objects on the plate. (2) For each of these stars, superimpose all of the objects at fixed distances from the primary objects.  If wedge objects are present, there will be a peak in the secondary object distrubition.  Two peaks indicate a grating plate. (3) If a Pickering Wedge plate identification has been made, a plot of the difference of instrumental magnitudes of both the wedge and primary objects against the instrumental magnitude of the primary object will show two distributions: normal stars and the wedge objects.  (4) Finally, go through the entire Sextractor population and flag as Pickering Wedge objects all objects which meet the above position and magnitude criteria.

\subsection{Star Matching, Defect Filter, Photometry Calibration, and Magnitude Calculation}

These steps are described in more detail in \citet{laycock2010}, \citet{tang2010}, Tang, S. \citetext{in preparation},  and on the DASCH website\footnote{see http://hea-www.harvard.edu/DASCH/photometry.php}.  After wedge filtering, objects are matched either to the GSC2.3.2 catalog \citep{lasker2008} or the Kepler Input Catalog (for calibration of the Kepler satellite field).  A defect filter removes emulsion defects, dust, and development defects by comparing PSF characteristics of matched objects with unmatched objects.  Next, the plate is divided into nine annular bins and a colorterm algorithm estimates the spectral response of the plate.  For each annular bin, a lowess curvefitting algorithm is used to generate a calibration between the Sextractor instrumental magnitudes and the blue catalog magnitudes.  Finally, a local correction algorithm is used on a 50 x 50 grid to account for variations in emulsion and/or sky conditions. 

\subsection{Multiple Exposure Loop}

The multiple exposure loop removes the objects successfully matched to the calibration catalog from the SExtractor source list and submits the remainder source list to astrometry.net.  At the completion of the WCS fitting procedures, the new WCS parameters are applied to all objects in the original SExtractor dataset and the pipeline proceeds normally from Pickering Wedge filtering to completion. 

Two ambiguities arise:  (1) If a catalog object from one solution coincides with a different catalog object from another solution, then the result is considered a ``Multiple Exposure Blend''.  (2) If an object is not matched to any catalog object, these objects receive a special flag because it is impossible to assign the object to a particular exposure.

To prevent infinite loops, the algorithm is terminated if a new solution is close to or overlaps a previous solution.
 
\subsection{Photometry Database}

Magnitude measurements are stored in a set of binary files optimized for performance. Supporting star-specific and plate-specific data appear in MySQL tables.  A detailed database design document will appear in a future publication.

\section{Results and Discussion}

Results from a sample of 87 multiple exposure plates with full logbook transcriptions show  successful detection of 31\% of the double exposures and 18\% of the triple exposures. There were 481 non-Pickering-Wedge plates which showed more solutions than described in the transcribed logbook entries.
  
The multiple exposure loop provided additional quality control checks by detecting at least 292 plates that can not provide good photometry because they are grating plates or plates with bad astrometric solutions .

The Pickering Wedge filter flagged 857 of 11461 scanned plates. Use of this filter has been extended to detect a possible optics misalignment in the ``dsy'' telescope series. These results suggest that a general near-neighborhood search algorithm would be useful for better detection of grating images and double-peaked star PSF's.

\acknowledgements This work is funded by National Science Foundation grants AST-0407380 and AST-0909073.  Additional contributers to the DASCH project are the Harvard College Plate Stacks Curator, Alison Doane; hardware engineer, Robert Simcoe; and our transcribers, plate cleaners, and scanners.
\bibliography{P033}

\end{document}